
\documentstyle{amsppt}
\document
\centerline{\bf Central extensions of current groups in two dimensions}

\vskip .20in
\centerline{\bf Pavel I. Etingof and Igor B. Frenkel}
\vskip .15in
\centerline{Yale University}
\centerline{Department of Mathematics}
\centerline{2155 Yale Station}
\centerline{New Haven, CT 06520 USA}
\centerline{e-mail etingof\@pascal.math.yale.edu}
\vskip .10in
\centerline{April 1992}
\centerline{Submitted to Communications in  Mathematical Physics in
February 1993}
\centerline{\bf Introduction}

\vskip .1in

The theory of loop groups and their representations \cite{13} has
recently developed in an extensive field with deep connections to many
areas of mathematics and theoretical physics. On the other hand, the
theory of current groups in higher dimensions contains rather isolated
results which have not revealed so far any deep structure comparable
to the one-dimensional case. In the present paper we investigate the
geometry of current groups in two dimensions and point out several
remarkable similarities with loop groups. We believe that these
observations give a few more hints about the existence of a new vast
structure in dimension two.

One of important problems in the theory of loop groups is integration
of central extensions of loop algebras. It is known \cite {13} that
the non-trivial one dimensional central extension of the loop algebra
${\frak g}^{S^1}$ of a compact Lie algebra $\frak g$ integrates to a Lie
group, which is a one dimensional central extension of the loop group
$G^{S^1}$ for the corresponding compact group $G$. Topologically
this group turns out to be a nontrivial circle bundle over $G^{S^1}$,
which plays a crucial role in the geometric realization
of representations of affine Lie algebras. The study of the coadjoint
action of this group yields a very simple description of orbits
by means of the theory of ordinary differential equations \cite{5,16}, and it
turns out that there is a perfect correspondence between orbits and
representations \cite {5}.

The Lie algebra of vector fields on the circle $\text{Vect}(S^1)$
arises as the Lie algebra of outer derivations of a loop algebra and
has many similarities with loop algebras. It has a unique nontrivial
one-dimensional central extension -- the Virasoro algebra. The structure
of coadjoint orbits of the Virasoro algebra can be obtained from the study of
Hill's operator\cite{7,16}. The invariants of the coadjoint action of the
complex Virasoro algebra are
essentially the same as
those of the extended algebra of loops in ${\frak sl_2}$.

In this paper we generalize some of these results for loop algebras and
groups
as well as for the Virasoro algebra to the two-dimensional case.

We define and study a class of infinite dimensional
complex Lie groups which are central
extensions of the group of smooth maps from a two dimensional
orientable surface without boundary to a simple complex Lie group
$G$.
These extensions naturally correspond to complex curves.
The kernel of such an extension is the Jacobian of the curve. The study of
the coadjoint action shows
that its orbits are labelled by moduli of holomorphic principal
$G$-bundles
over the curve and can be described in the
language of partial differential equations. In genus one it is also
possible to describe the orbits as conjugacy classes of
the twisted loop group, which leads to consideration of difference
equations for holomorphic functions. This gives rise to a hope
that the described groups should possess a counterpart of the
rich representation theory that has been developed for loop groups.

We also define a two-dimensional analogue of the Virasoro algebra
associated with a complex curve. In genus one, a study of a complex
analogue of Hill's operator yields a description of invariants of
the coadjoint action of this Lie algebra. The answer turns out to be
the same as in dimension one: the invariants coincide with those
for the extended algebra of currents in $\frak sl_2$.

Note that our main constructions are purely two-dimensional. In
particular, in three and more dimensions the space of orbits
tends to be infinite-dimensional and topologically unsatisfactory.

\heading
{\bf 1. Lie algebra extensions.}
\endheading

Let $G$ be a simply connected simple complex Lie group, and let
$\frak g$ be its Lie algebra. Denote by $<,>$ a nonzero invariant bilinear
form on $\frak g$. Let $\Sigma$ be a nonsingular two-dimensional surface of
genus $g$. Define $G^{\Sigma}$ as the group of
all smooth maps from $\Sigma$ to $G$. Its Lie algebra consists of
all smooth maps from $\Sigma$ to $\frak g$ and will be denoted by
$\frak g^{\Sigma}$. These are called \it the current group \rm and
\it the current algebra \rm.

Now fix a complex structure on the surface $\Sigma$.
Let $H_{\Sigma}$  be the space of holomorphic
differentials on $\Sigma$. Its dimension is equal to $g$. Let $\omega$
be the identity element in $H_{\Sigma}\otimes H^*_{\Sigma}$. The
element $\omega$
can be regarded as a holomorphic differential on $\Sigma$ with values
in  $H^*_{\Sigma}$. Define a 2-cocycle on $\frak g^{\Sigma}$ with
values in $H^*_{\Sigma}$ (regarded as a trivial $\frak g^{\Sigma}$-module) by
$$
\Omega(X,Y)=\int_{\Sigma}\omega\wedge<X,dY>,\quad X,Y\in \frak
g^{\Sigma}.\tag 1.1
$$
This cocycle defines a $g$-dimensional central extension of
$\frak g^{\Sigma}$. This extension is non-trivial for surfaces of
positive genus, and it has no
nonzero trivial subextensions or quotient extensions.
Denote this new Lie algebra by $\hat{\frak g}^{\Sigma}$.

If $\Sigma$ is a complex torus, we can separate a subalgebra
$\hat{\frak g}^{\Sigma}_{pol}$ in
$\hat{\frak g}^{\Sigma}$ that consists of all currents realized by
trigonometric polynomials. The restriction of the above central
extension to this subalgebra in its natural basis looks very simple.
Namely, let $L$ be a lattice in the complex plane
$\Bbb C$ such that $\Sigma=\Bbb
C/L$. The Lie algebra $\hat{\frak g}^{\Sigma}_{pol}$ is
spanned by elements $x(n)$, $x\in\frak g$, $n\in L$, where $x(n)$
depends linearly on $x$ for any fixed $n$. These elements satisfy the
following commutation relations:
$$
[x(n),y(m)]=[x,y](n+m)+n\delta_{n,-m}<x,y>k\tag 1.2
$$
where k is the central element.
This extension is a natural two-dimensional counterpart of affine
algebras. Similar extensions were considered in \cite{10}.

Each of the extensions defined above can be obtained as a suitable
quotient of the universal central extension $U\frak g^{\Sigma}$
of the Lie algebra $\frak g^{\Sigma}$.

\proclaim{Proposition 1.1} (\cite{13, Section 4.2}) The universal central
extension $U\frak g^{\Sigma}$ is an extension of $\frak g^{\Sigma}$ by
means of the infinite-dimensional space ${\frak a}=\Omega^1(\Sigma)/d
\Omega^0(\Sigma)$ of complex-valued
1-forms on $\Sigma$ modulo exact forms. This extension is defined by the
${\frak a}$-valued cocycle
$$
u(\xi,\eta)=<\tilde\xi,d\tilde\eta>,\quad \xi,\eta\in {\frak a},
\tag 1.3
$$
where $\tilde\xi, \tilde\eta\in \Omega^1(\Sigma)$ are any liftings of
$\xi, \eta$.
\endproclaim

Note that this construction does not involve the complex structure on
the surface.

In order to obtain $\hat{\frak g}^{\Sigma}$ for a specific complex structure
on the surface $\Sigma$ as a quotient of
$U\frak g^{\Sigma}$, one needs to factorize it by the subgroup
of all $\xi\in {\frak a}$ such that
$$
\int_{\Sigma}\tilde\xi\wedge \omega=0\tag 1.4
$$
for any lifting $\tilde\xi\in \Omega^1(\Sigma)$ of the element $\xi$.

\proclaim{Proposition 1.2}
(i)  Let $\Sigma_1$ and $\Sigma_2$ be two Riemann surfaces. The Lie algebras
$\hat{\frak g}^{\Sigma_1}$ and $\hat{\frak g}^{\Sigma_2}$ are
isomorphic if and only if $\Sigma_1$ and $\Sigma_2$ are conformally
equivalent.

(ii)  Any automorphism $f$ of $\hat{\frak g}^{\Sigma}$
can be uniquely represented as a composition: $f=h\circ \phi_*$ where
$h$ is a conjugation by an element of
$\text{Aut}(\frak g)^{\Sigma}$
and $\phi_*$ is the direct image map induced by a conformal
diffeomorphism $\phi: \Sigma\to\Sigma$.
\endproclaim

\demo{Proof} Let $f:$ $\hat{\frak g}^{\Sigma_1}\to\hat{\frak g}^{\Sigma_2}$
 be any isomorphism. Then the dimensions of the centers of
$\hat{\frak g}^{\Sigma_1}$ and $\hat{\frak g}^{\Sigma_2}$ must coincide,
so the genera of $\Sigma_1$ and $\Sigma_2$ are the same. Consider the
induced map $f_0:$ ${\frak g}^{\Sigma_1}\to{\frak g}^{\Sigma_1}$
of current algebras. Since $\Sigma_1$ and $\Sigma_2$ are
diffeomorphic, we may actually regard this map as an automorphism of
the current algebra. According to \cite {13, Section 3.4}, any such map
uniquely decomposes as $f_0=h\circ \phi_*$
where
$h$ is a conjugation by an element of $\text{Aut}(\frak g)^{\Sigma_1}$
and $\phi_*$ is the direct image map induced by a conformal
diffeomorphism $\phi: \Sigma_1\to\Sigma_2$. In order for $f_0$ to
extend to an isomorphism of extensions, $\phi$ must take holomorphic
differentials to holomorphic differentials which forces it to be a
conformal equivalence. This proves both (i) and (ii).$\blacksquare$
\enddemo

One can also classify embeddings $\hat{\frak
g}^{\Sigma_1}\hookrightarrow\hat{\frak g}^{\Sigma_2}$. Indeed, any
embedding $f:$ ${\frak
g}^{\Sigma_1}\hookrightarrow{\frak g}^{\Sigma_2}$
is induced by a surjective map
$\varphi:$ $\Sigma_2\to\Sigma_1$, and conversely, any such map defines an
embedding.
In order for this embedding to continue to central extensions, the map
between surfaces must be holomorphic. However, in this case the
continuation is not unique. Continuations correspond to
monomorphisms $\chi: H_{\Sigma_1}\hookrightarrow H_{\Sigma_2}$ with the
property $\varphi_*(\chi(v))=v$ for all $v\in H_{\Sigma_1}$. Here
$\varphi_*$ is the
map $H_{\Sigma_2}\to H_{\Sigma_1}$ induced by $\varphi$.

We have shown that the group of outer automorphisms of the Lie
algebra $\hat{\frak g}^{\Sigma}$ is isomorphic to the group of
holomorphic automorphisms of the Riemann surface $\Sigma$. If $g>1$, this group
is finite, and it is trivial for almost every
surface. In spite of it, $\hat{\frak g}^{\Sigma}$ has plenty of
outer derivations.

\proclaim{Proposition 1.3}
If $g>1$, the Lie algebra of outer derivations of $\hat{\frak g}^{\Sigma}$
coincides with the Lie algebra $\text{Vect}_{0,1}(\Sigma)$ of all
complex-valued
vector fields on $\Sigma$ of
type (0,1), i.e. of the form $u(z,\bar{z})\frac{\partial}{\partial\bar
z}$ for any local complex coordinate $z$, $u$ being a smooth function.

If $g=1$, the Lie algebra of outer derivations is
$
<\frac{\partial}{\partial z}>\ltimes \text{Vect}_{0,1}(\Sigma)$
\endproclaim

\demo{Proof} It is known that outer derivations of
${\frak g}^{\Sigma}$
are in one-to-one correspondence with complex vector fields on the
surface $\Sigma$. In order for such a derivation to continue to the
central extension $\hat{\frak g}^{\Sigma}$, the vector field must
annihilate holomorphic differentials. If $g>1$, such a field must have type
(0,1). If $g=1$, any such field is a linear combination of a
(0,1)-field and the constant field
$\frac{\partial}{\partial
z}$ $\blacksquare$
\enddemo

The Lie algebra $\text{Vect}_{0,1}(\Sigma)$ should be thought of as an
analogue of the Witt algebra. It has a $g$-dimensional central
extension which is a natural analogue of the Virasoro algebra.
This is the extension by the space $H_{\Sigma}$ defined by the cocycle
$$
F(X,Y)=\int_{\Sigma}\omega\wedge\bar{\partial}X\bar{\partial}^2Y,
X,Y\in \text{Vect}_{0,1}(\Sigma).\tag 1.5
$$
In this expression, $\bar{\partial}X$ is a function and
$\bar{\partial}^2Y$ is a differential 1-form, since the operator
$\bar{\partial}$ maps vector fields to functions and functions to 1-forms.
We will denote this extension by $\text{Vir}(\Sigma)$.
For $\Sigma$ being the torus, the polynomial part
$\text{Vect}_{0,1}(\Sigma)
_{\text{pol}}$ of
$\text{Vect}_{0,1}(\Sigma)$
is a graded Lie algebra of a very simple structure. Namely, let $L$ be
the lattice in $\Bbb C$ such that $\Sigma=\Bbb C/L$. The basis
of $\text{Vect}_{0,1}(\Sigma)_{\text{pol}}$
consists of elements $e_n$, $n\in L$, which satisfy Witt's relations
$$
[e_n,e_m]=(m-n)e_{n+m}.\tag 1.6
$$
The one dimensional extension we have described has an additional
basis vector $c$ -- the central element, and the relations are
analogous to Virasoro relations:
$$
[e_n,e_m]=(m-n)e_{n+m}+m^3\delta_{m,-n}c.\tag 1.7
$$
One can easily show that this is the universal central extension
of $\text{Vect}_{0,1}(\Sigma)_{\text{pol}}$ (i.e. it is the only possible
nontrivial one-dimensional extension up to an isomorphism). The proof
is similar to that for the Witt algebra and is given in \cite {15}.

\heading
{\bf 2. Group extensions.}
\endheading

The following theorem describes the topology of the current group and
can be deduced from the results of \cite {13, Chapter 4}.

\proclaim{Theorem 2.1}

(i) $G^{\Sigma}$ is connected.

(ii) $\pi_1(G^{\Sigma})=\Bbb Z$.

(iii) $\pi_2(G^{\Sigma})\otimes \Bbb R=H_2(G^{\Sigma}, \Bbb R)=\Bbb
R^{2g}$.

\endproclaim

It turns out that the Lie algebra $\hat{\frak g}^{\Sigma}$ can be
integrated to a complex Lie group.

\proclaim{Theorem 2.2} There exists a central extension $\hat G^{\Sigma}$
of the current group $G^{\Sigma}$ by means of the Jacobian variety of $\Sigma$
whose Lie algebra is $\hat{\frak g}^{\Sigma}$.
\endproclaim
\demo{Proof}
This group can be constructed by the
procedure described in \cite {13, Chapter 4}. Namely, consider
the left invariant holomorphic 2-form on $G^{\Sigma}$ equal to
$\Omega$ on the tangent space at the identity. It shall be denoted by the
same letter. Integrals of $\Omega$ over integer 2-cycles in
$G^{\Sigma}$ fill the lattice $L=H_1(\Sigma,\Bbb Z)$ in
$H^*_{\Sigma}$. Therefore, there exists a holomorphic principal
bundle over $G^{\Sigma}$ with fiber $J=H^*_{\Sigma}/L$
and with a holomorphic connection $\theta$ whose curvature form is
$2\pi\Omega$. Note that
$J$ is the Jacobian of the surface $\Sigma$. Define the group $\hat
G^{\Sigma}$
as the group of all transformations of the constructed bundle that
preserve the connection $\theta$ and project to left translations on
the group $G^{\Sigma}$. It is a central extension of $G^{\Sigma}$
by $J$. One has an exact sequence
$$
1\to J\to \hat G^{\Sigma}\to G^{\Sigma}\to 1,\tag 2.1
$$
and it follows that the Lie algebra of $\hat G^{\Sigma}$ is isomorphic
to $\frak g^{\Sigma}$. $\blacksquare$
\enddemo

Obviously, this theorem will remain valid if we replace the current
group $G^{\Sigma}$ with its universal covering $\tilde G^{\Sigma}$.
Denote the extension obtained in this way by $\widetilde{\hat G}^{\Sigma}$.

The following theorem characterizes the universal central extension of
the current group.

\proclaim{Theorem 2.3}\cite {13, Section 4.10}
(i) The universal central extension $UG^{\Sigma}$ of the group
$G^{\Sigma}$ is
 an extension of $\tilde G^{\Sigma}$
by means of the infinite-dimensional abelian
group $\Cal A$ of complex-valued 1-forms on $\Sigma$ modulo
closed 1-forms with integer periods.

(ii)
$\pi_n(UG^{\Sigma})\otimes \Bbb R=0$ for $n<3$.

(iii) The group $\Cal A$ is homotopy equivalent to a $2g$-dimensional torus.
The natural map $\pi_2(G^{\Sigma})\to\pi_1(\Cal A)$ associated to the fiber
bundle $UG^{\Sigma}\to G^{\Sigma}$ is an isomorphism up to torsion.

\endproclaim

\proclaim{Proposition 2.4}
The universal central extension
$UG^{\Sigma}$ is homotopy equivalent to $\widetilde{\hat G}^{\Sigma}$.
\endproclaim

\demo{Proof} Consider a homomorphism $\varepsilon:\Cal A\to J$ defined
by
$$
\varepsilon(a)=\int_{\Sigma}\tilde a\wedge\omega\quad \text{mod }
L.\tag 2.2
$$
where $\tilde a$ is any lifting of $a\in\Cal A$ into the space of
1-forms on $\Sigma$. It is easy to check that this map is well defined,
i.e. independent of the choice of $\tilde a$. Obviously, $\varepsilon$
is a homotopy equivalence. This implies that the corresponding map of
extensions $\tilde\varepsilon:UG^{\Sigma}\to \widetilde{\hat G}^{\Sigma}$
of the group $\tilde G^{\Sigma}$ is also a homotopy equivalence.
\enddemo

This theorem implies that $\widetilde{\hat G}^{\Sigma}$ is a nontrivial
$J$-bundle on $\tilde G^{\Sigma}$ and represents the homotopically
nontrivial part of the universal central extension.

In fact, the universal extension can be constructed rather explicitly
 as a central extension of $\tilde{ G^{\Sigma}}$ by the additive group
of a vector space, as follows.

Let $K$ be the maximal compact subgroup of $G$, and let
$\tilde{K}^{\Sigma}$ be the subgroup of all elements in $\tilde{
G^{\Sigma}}$
whose projections to ${ G^{\Sigma}}$ are $K$-valued currents.
Obviously, $\tilde{ K^{\Sigma}}$
 is a central extension of $K^{\Sigma}$ by $T^{2g}\times\Bbb Z$,
where $T^{2g}$ is the real $2g$-dimensional torus.

Choose a metric on $\Sigma$ compatible to the complex structure.
A metric induces an inner product on the space of differential forms.
Define $d^*$ to be the conjugate operator to the de Rham differential
$d$. This operator maps 2-forms to 1-forms. Let ${\frak a}$ be the space of
1-forms on $\Sigma$ modulo exact 1-forms, and let $W$ be the image
of the projection $p:$  $\text{Im}d^*\to {\frak a}$.

\proclaim{Lemma 2.5}
The subspace $W$ has
codimension $2g$ in ${\frak a}$ and is complementary to the subspace
$H^1(\Sigma,\Bbb
R)\subset {\frak a}$.
\endproclaim

\demo{Proof} Let $\Delta:\Omega^1(\Sigma)\to\Omega^1(\Sigma)$ be the
Laplacian associated with the metric: $\Delta=dd^*+d^*d$.
Since $\Delta$ is self-adjoint,
any form $\alpha\in \Omega^1(\Sigma)$ can be represented in the form
$\alpha=\alpha_1+\alpha_2$, where $\alpha_1\in \text{Ker}\Delta$,
$\alpha_2\in\text{Im}\Delta$. Then $d\alpha_1=0$ by Hodge's theorem,
and $\alpha_2=\Delta\beta=d^*d\beta+dd^*\beta=d^*\beta_1+d\beta_2$ by
construction. Thus, $p(\alpha)=p(\alpha_1)+p(d^*\beta_1)$. Since
$p(\alpha_1)\in H^1(\Sigma,\Bbb R)$ and $p(d^*\beta_1)\in W$, we have
shown that any vector in $\frak a$ can be written as a sum of a
vector in $W$ and a vector in $H^1$.

Now let us show that $W\cap H^1(\Sigma,\Bbb R)=0$. Let $w$ be an
element of this intersection. Let $\tilde w$ be an inverse image of
$w$ in $\text{Im}d^*$. Then $d\tilde w=0$ and $d^*\tilde w=0$, so
$w$ is harmonic. By Hodge's theorem, harmonic forms are in one-to-one
correspondence with cohomology classes, so any harmonic form in
$\text{Im}d^*$ has to be zero. Thus, $w=0$.$\blacksquare$
\enddemo

Define a 2-cocycle on $\tilde{ K^{\Sigma}}$ with values in $W$ as
follows:
$$
C(\tilde{a},\tilde{b})=p(d^*<b^{-1}db,da\cdot a^{-1}>),\quad \tilde{a},
\tilde{b}\in \tilde{ K^{\Sigma}}\tag 2.3
$$
where $a,b\in { K^{\Sigma}}$
are images of $\tilde{a},\tilde{b}$.

This cocycle defines a central extension of $\tilde{ K^{\Sigma}}$ by
$W$. Let us show that this new group is isomorphic to the universal central
extension of $K^{\Sigma}$.

Since both groups are simply connected, it is enough to establish
an isomorphism between their Lie algebras. Both Lie algebras are
central extensions
of $\frak k^{\Sigma}$, where $\frak k$ is the Lie algebra of $K$. The
kernels of these extensions are $\tilde{\frak a}=W\oplus H_{\Sigma
\Bbb R}$ and $\frak a$,
respectively, and the corresponding cocycles are $\tilde u=C\oplus\Omega_{\Bbb
R}$ and $u$ (if $V$ is a vector space then $V_{\Bbb R}$ denotes the
space $V$
regarded as a real vector space).
All we need is to construct an isomorphism $\rho:{\frak a}\to
\tilde{\frak a}$ such that
$\rho(u)=\tilde u$. It is easy to check that such an isomorphism is provided
by Lemma 2.5.

The universal extension of $\tilde{ G^{\Sigma}}$ can now be obtained
from the constructed extension of $K^{\Sigma}$ by complexification.

\proclaim{Remark 2.1}\rm
 (0,1)- vector fields on the surface are realized as
$J$-invariant holomorphic vector fields on $\hat G^{\Sigma}$ (=first order
differential operators on holomorphic functions). Any such
field $v$
satisfies the condition $v(gh)=v(g)h+gv(h)$ for all $g,h\in\hat G^{\Sigma}$.
\endproclaim

\proclaim{Remark 2.2}\rm
A very challenging problem is to obtain a more explicit construction of
$\widetilde{\hat G}^{\Sigma}$ than we have given, by presenting a 2-dimensional
counterpart of the famous Wess-Zumino-Witten construction for loop
groups \cite {19,9}.

It would also be interesting to obtain a direct construction of the
universal central extension of $G^{\Sigma}$ from $\widetilde{\hat G}^{\Sigma}$
by an analogue of formula (2.3) without referring to the
maximal compact subgroup.
\endproclaim

\heading
{\bf 3. Orbits of the coadjoint action.}
\endheading

Let $\eta$ be a holomorphic differential on $\Sigma$. Denote by
$E_{\eta}$ the one-dimensional central extension of $\frak g^{\Sigma}$
by means of the cocycle
$$
\Omega_\eta(X,Y)=\int_{\Sigma}\eta\wedge<X,dY>.\tag 3.1
$$
Obviously, $E_{\eta}$ is the quotient of $\hat{\frak g}^{\Sigma}$ by
the kernel of $\eta$ as a linear function on $H_{\Sigma}$.

Denote by $E^*$ the space of all operators
$D=\lambda\bar{\partial}+\xi$, where $\lambda\in\Bbb C$ and $\xi$
is a $\frak g$-valued (0,1) form on $\Sigma$, i.e. a 1-form which
can be written as $u(z,\bar{z})d\bar{z}$ for any local complex coordinate
$z$, $u$ being a $\frak g$-valued smooth function.
For any representation $V$ of $\frak g$, these operators take $V$-valued
functions on $\Sigma$ to $V$-valued (0,1)-forms according to
$$
D\psi=\lambda\bar{\partial}\psi+\xi\psi,
$$
which allows to define an action of $G^{\Sigma}$ on $E^*$:
$$
h\circ (\lambda\bar{\partial}+\xi)=\lambda\bar{\partial}+h^{-1}
\bar{\partial} h+\text{Ad}h\circ \xi,\quad h\in G^{\Sigma}.\tag 3.2
$$

Consider a pairing between $E^*$ and $E_{\eta}$ given by
$$
(\lambda\bar{\partial}+\xi,\mu
k+X)=\lambda\mu+\int_{\Sigma}\eta\wedge
<\xi,X>,\tag 3.3
$$
where $k$ is the basis central element in $E_{\eta}$ and $\mu\in\Bbb C$.

\proclaim{Proposition 3.1} Pairing (3.3) is $G^{\Sigma}$-invariant and
nondegenerate.
\endproclaim

The proof is straightforward.

The constructed pairing allows us to interpret $E^*$ as the smooth
part of the dual $E_{\eta}^{\prime}$ to the Lie algebra $E_{\eta}$,
i.e. as the proper coadjoint representation.

Our goal now is to study the orbits of the action of $G^{\Sigma}$ in
$E^*$. Hyperplanes $\lambda=\text{const}$ are invariant under this
action. Let us fix a nonzero value of
$\lambda$ and examine the orbits contained in the corresponding
hyperplane $\Cal H_{\lambda}$.

We will use the following classical construction \cite {4}.

To an operator $D=\lambda\bar{\partial}+\xi$ one can naturally
associate
a holomorphic principal $G$-bundle on $\Sigma$ as follows.
Let $U_i,i\in I,$ be a cover of $\Sigma$ by proper open subsets.
For every $i$ fix a solution $\psi_i:U_i\to G$ of the partial
differential equation
$$
\lambda\bar{\partial}\psi_i+\xi\psi_i=0.\tag 3.4
$$
(the left hand side of (3.4) being a $\frak g$-valued (0,1) form on
$U_i$). Such solutions can always be found.
Now for $i,j\in I$ define transition functions $\phi_{ij}:U_i\cap
U_j\to G$ by $\phi_{ij}=\psi_i^{-1}\psi_j$. It is easy to see that
these functions are holomorphic. Thus they define a holomorphic
principal $G$-bundle on $\Sigma$, which we denote by $B(D)$. Moreover,

it follows that the
holomorphic bundles $B(D_1)$ and $B(D_2)$ associated to two operators $D_1
=\lambda\bar{\partial}+\xi_1$ and $D_2=\lambda\bar{\partial}+\xi_2$
are equivalent if and only if there exists an element $h\in
G^{\Sigma}$ such that $h\circ D_1=D_2$. This $h$ will be exactly the
gauge transformation establishing the equivalence between $B(D_1)$ and
$B(D_2)$.

Conversely, for any holomorphic principal $G$-bundle $B$ on $\Sigma$
there exists an operator $D=\lambda\bar{\partial}+\xi$ such that
$B=B(D)$. Indeed, any principal $G$-bundle on $\Sigma$ is
topologically trivial, since $G$ is simply connected. If we choose a global
trivialization then the local holomorphic trivializations over open
sets $U_i$ will be expressed
by smooth functions $\psi_i:U_i\to G$. Since the transition functions
$\phi_{ij}=\psi_i^{-1}\psi_j$
are holomorphic on $U_i\cap U_j$, we have $\bar{\partial}\psi_i\cdot
\psi_i^{-1}=\bar{\partial}\psi_j\cdot
\psi_j^{-1}$ on $U_i\cap U_j$. Therefore there exists a $\frak
g$-valued 1-form $\xi$ on $\Sigma$ such that
$\xi=-\lambda\bar{\partial}
\psi_i\cdot
\psi_i^{-1}$ on $U_i$ for all $i\in I$. Let
$D=\lambda\bar{\partial}+\xi$. Then $B=B(D)$.

This reasoning proves the following
\proclaim{Proposition 3.2}
Orbits of the action of $G^{\Sigma}$ in $\Cal H_{\lambda}$
are in one-to-one correspondence with equivalence classes of
holomorphic $G$-bundles on $\Sigma$. The correspondence
is $D\leftrightarrow B(D)$.
\endproclaim
\proclaim{Remark 3.1}\rm
The relation between differential operators and holomorphic principal
bundles on $\Sigma$ was used in computations of partition functions of the
gauged Wess-Zumino-Witten model in \cite{6} and \cite{20}.
\endproclaim

\proclaim{Remark 3.2} \rm
Holomorphic sections of the bundle $B(D)$, i.e. solutions of (3.4),
are known as a special case of generalized analytic functions which
were introduced in the fifties by L.Bers and I.Vekua \cite {2,18}.
If $\xi=0$, they become usual analytic functions.
\endproclaim

Holomorphic principal $G$-bundles for $G=SL_n(\Bbb C)$
were classified by Atiyah \cite {1} for $g=1$ and by Narasimhan
and Sheshadri \cite {11,12}, for $g>1$. Their results were generalized
to the case of any simple group $G$ by Ramanathan \cite {14}. We summarize here
some of them.

Let $\Pi$ be the fundamental group of $\Sigma$ and $K$ be a maximal
compact subgroup of $G$. Let $\rho:\Pi\to
K$ be any homomorphism. This homomorphism defines a flat $G$-bundle
$B_{\rho}$ over $\Sigma$ with a canonical holomorphic structure.
Bundles coming from this construction are called unitary.
It is known \cite {14} that the conjugacy class of the representation
$\rho$ is completely determined by the equivalence class of
$B_{\rho}$ as a holomorphic principal bundle.

The following theorem shows that almost all
holomorphic principal bundles are flat and unitary.

\proclaim{Theorem 3.3}\rm\cite {14}
Let $\lbrace B_t\rbrace_{t\in \Cal T}$ be a holomorphic family of
holomorphic principal $G$-bundles parametrized by a complex space
$\Cal T$.
Then the  subset $\Cal T_0$ of such $t\in
\Cal T$ that the bundle $B_t$ is flat and unitary is Zariski open in
$\Cal T$.
\endproclaim

Let us now define an appropriate notion of ``almost everywhere''.

Let $V$ be any topological vector
 space. We say that a set $Z\subset V$ is Zariski
open if for any finite dimensional complex manifold $\Cal T$
and any holomorphic map $f:\Cal
T\to V$ the set of points $t\in \Cal T$ such that $f(t)\in Z$ is
Zariski open in $\Cal T$. Obviously, this defines a topology in $V$. We
will say that some property holds almost everywhere in $V$ if it holds
on a nonempty Zariski open subset of $V$. Note that every nonempty
Zariski open subset in $V$ is dense, open, and connected in the usual topology.

It is clear that a flat bundle $B$ can be obtained from an
operator $D=\lambda\bar{\partial}+\xi$ with $\xi$ being a $\frak g$-valued
antiholomorphic differential on $\Sigma$: $B=B(D)$. Therefore, the
above theorem in fact states that almost every (in Zariski sense)
differential operator
of the form $\lambda\bar{\partial}+\xi$ can be reduced to an operator
with an antiholomorphic $\xi$ by means of a gauge transformation from
$G^{\Sigma}$. This implies that the union $S$ of $G^{\Sigma}$-orbits
of all operators $\lambda\bar{\partial}+\xi$ with $\xi$ being
antiholomorphic is a Zariski open subset of $\Cal H_{\lambda}$.

\proclaim{Remark 3.3}\rm  It is easy to construct an example of a holomorphic
bundle which is not flat and unitary. For instance, take any holomorphic line
bundle $\beta$ on $\Sigma$ of degree 1 and form a rank 2 bundle
$\beta\oplus\beta^*$, where $\beta^*$ is the dual bundle to $\beta$.
This bundle has degree 0. Let $B=\text{Aut}_0(\beta\oplus\beta^*)$ be the
$SL_2(\Bbb C)$-bundle of automorphisms of
$\beta\oplus\beta^*$ having determinant one. It is easy to see that
this bundle is not flat: the associated bundle $\beta\oplus\beta^*$
has a nonzero holomorphic section $s$ which vanishes at a point $z\in
\Sigma$; this section cannot satisfy any equation $\lambda\bar{\partial}s+\xi
s=0$ with $\xi$ being antiholomorphic. Another example would be
$\text {Aut}_0(\Cal A)$ where $\Cal A$
is the Atiyah's bundle of rank 2 over a complex torus which is
a semidirect sum of two trivial line bundles. This bundle is flat, but
not unitary. However, according
to the above theorem,
these bundles are exceptional and can be made flat and
unitary by an arbitrarily small perturbation of
transition functions.
\endproclaim

The space of equivalence classes of flat and unitary $G$-bundles is a
complex manifold with singularities. Denote this manifold by $\Cal M$.
This manifold can be thought of as the space of coadjoint orbits of
generic position.

We can now classify all holomorphic invariants of the action of
$\frak g^{\Sigma}$ in $\Cal H_{\lambda}$.

Let $\frak b$ be a complex Lie algebra and $V$ be a topological
 representation of
$\frak b$. Define a holomorphic invariant of the action of $\frak b$
in $V$ as a holomorphic map from a Zariski open subset in $V$ to a
fixed
finite-dimensional
complex manifold which is invariant under the action of $\frak b$.
The coefficient $\lambda$ is an example of a holomorphic
invariant of the coadjoint action of $\frak
g^{\Sigma}$. Another example would be the $\Cal M$-valued function
 $b$ defined on a Zariski open subset in $E^*$ which
takes an operator $D$ to the equivalence class of $B(D)$.
The orbit classification result implies

\proclaim{Proposition 3.4} Any holomorphic invariant of the action of
$\frak
g^{\Sigma}$ in $E^*$ is a function of $b$ and $\lambda$.
\endproclaim

\heading
{\bf 4. Orbital structure in genus 1.}
\endheading

In this section we assume the surface $\Sigma$ to be a complex torus
$\Bbb C/L$, $L=\lbrace p+q\tau|p,q\in\Bbb Z\rbrace$, $\text{Im}\tau>0$.
In this case it is possible to give a more explicit description of the
space of orbits.

The coadjoint representation $E^*$ of the group $\hat G^{\Sigma}$ can
be identified with the space of differential operators
$\lambda\frac{\partial}{\partial\bar z}+\xi$ where $\xi$ is a $\frak
g$-valued function on the torus. These operators act on $V$-valued
functions on the torus for any representation $V$ of $G$. According to
the previous section, orbits of the
coadjoint action of $G^{\Sigma}$ restricted to a hyperplane $\Cal
H_{\lambda}$ correspond to equivalence classes
of holomorphic principal $G$-bundles over the complex torus $\Sigma$.
Almost every such bundle is flat and unitary, i.e. comes from a
homomorphism from the fundamental group to the maximal compact subgroup $K$
in $G$.
The fundamental group of the torus is $\Pi=\Bbb Z^2$, so we may assume
that this homomorphism lands in a maximal torus $T\subset K$. Since
all maximal tori are conjugate, we can assume $T$ to be a fixed
maximal torus in $K$.
The images of the two generators of $\Pi$ can be any two elements in
$T$. Thus, we get a covering of the set of non-equivalent unitary
representations of
$\Pi$ by the product $T\times T$. Two elements of this product
correspond to isomorphic representations of $\Pi$ if and only if
one of them can be obtained from the other by the action of an element
of the Weyl group $W$ of $K$. Therefore, the set of equivalence
classes of unitary representations of $\Pi$ is $(T\times T)/W$.

Thus, we have found that topologically the space $\Cal M$ of orbits of
generic position is $(T\times T)/W$. However, this realization does
not tell us anything about the complex structure on $\Cal M$.
Therefore let us describe a different realization of $\Cal M$.

Let $D=\lambda\frac{\partial}{\partial\bar z}+\xi$. If the bundle
$B(D)$ is flat (which happens, as we know, for almost every $\xi$)
then the equation $D\psi=0$ on a $G$-valued function $\psi$
will have a solution $\psi(z)$ with the properties
$\psi(z+1)=\psi(z)$, $\psi(z+\tau)=\psi(z)A$, where $A$ is a fixed
element of $G$. If the bundle $B(D)$ is also unitary, the element $A$
will be semisimple. Then we may assume that it belongs to a fixed
complex maximal torus $T_{\Bbb C}\subset G$.

The element $A$ completely determines the bundle $B(D)$. Indeed, let
$a\in \frak g$ satisfy $A=\exp(a)$, and set
$F(z)=\psi(z)\exp(-a\frac{z-\bar z}{\tau-\bar{\tau}})$. Obviously
$F$ is doubly periodic, i.e. $F\in G^{\Sigma}$, and we have
$$
F\circ(\lambda\frac{\partial}{\partial\bar
z}+\xi)=\lambda\frac{\partial}{\partial\bar
z}-\frac{a}{\tau-\bar{\tau}}.\tag 4.1
$$

Moreover, different elements of $T_{\Bbb C}$ may correspond to
equivalent bundles. Let $r$ be the rank of $G$ and $\frak t$ be the
Lie algebra of $T$. If
$h_j\in i\frak t$, $1\le j\le r$ denote the standard basis of the Cartan
subalgebra of $\frak g$ then $\exp(2\pi ih_j)=1$, and
the solution $\psi(z)$ of the equation $D\psi=0$ can be replaced by
another solution $\psi_n(z)=\psi(z)\exp(2\pi iz(\sum_{j=1}^{r}n_jh_j))$ where
$n_j\in \Bbb Z$. This solution satisfies $\psi_n(z+1)=\psi_n(z)$,
$\psi_n(z+\tau)=\psi(z)A_n$, where $A_n=A\exp(2\pi
i\tau(\sum_{j=1}^{r}n_jh_j))$. Obviously, $A_n$ corresponds to the
same bundle as $A$. Furthermore, if we conjugate $A$ by any element of
the Weyl group W, we will obtain an element $A^{\prime}$ of $T_{\Bbb C}$ that
corresponds to the same bundle as $A$.

Let $Q^{\vee}$ be the lattice generated by $h_j$ (the dual weight
 lattice of $G$).
We have shown that the bundle $B(D)$ is completely determined by the
projection of $A$ into the complex space $T_{\Bbb C}/(W\ltimes
\exp(2\pi i\tau Q^{\vee}))={\frak t}_{\Bbb C}/(W\ltimes (Q^{\vee}\oplus\tau
Q^{\vee}))$.
It is easy to see that different points of this space correspond to
non-isomorphic bundles. Therefore, we have

\proclaim{Proposition 4.1}
The space $\Cal M$ of equivalence classes of flat and unitary
holomorphic $G$-bundles
over the complex torus is isomorphic to ${\frak t}_{\Bbb C}/(W\ltimes
(Q^{\vee}\oplus\tau Q^{\vee}))$.
\endproclaim

An alternative way to obtain the classification of
 generic coadjoint orbits of $\hat
G^{\Sigma}$  is based on a
study of conjugacy classes of the twisted loop group. Let us give a brief
description of this method.

By
a loop group $G^{S^1}_h$ we mean the group of holomorphic maps from the
cylinder $\Bbb C/\Bbb Z$ to $G$. The abelian group $\Bbb C/\Bbb Z$
acts by automorphisms on $G^{S^1}_h$ through translations of the argument.
Form the semidirect product $\check G^{S^1}_h=\Bbb C/\Bbb Z\ltimes G^{S^1}_h$
associated to this action. Let us analyze the conjugacy classes of
$\check G^{S^1}_h$.  For $(z,f)$, $(\tau ,g)$ $\in \check G^{S^1}_h$ we have
$$
(z,f)^{-1}(\tau,g)(z,f)=(\tau,h),\quad h(u)=f(u-\tau)^{-1}g(u-z)f(u),\quad
u\in\Bbb C/\Bbb Z.\tag 4.2
$$
This implies that $\tau$ is preserved under conjugations, so we may
assume that it is fixed, and study conjugacy classes inside the coset
$C_{\tau}=\lbrace (\tau,g)|g\in G^{S^1}_h\rbrace$. We will treat the
case $\tau
\notin
\Bbb R/\Bbb Z$. In this case we may assume, without loss of
generality, that $\text{Im}\tau>0$.

Let $\xi$ be a smooth function on the torus $\Sigma_{\tau}$.
Consider the differential equation
$$
\lambda\frac{\partial\psi}{\partial\bar
z}+\xi\psi=0\tag 4.3
$$
with respect to a $G$-valued function $\psi$ on the cylinder $\Bbb
C/\Bbb Z$.

Let $\psi_0(z)$ be a solution of equation (4.3). Then $\psi_0(z+\tau)$
is
also a
solution, since the equation is invariant under a translation by
$\tau$. Therefore, $\nu_0(z)=\psi_0(z)^{-1}\psi_0(z+\tau)$ is a holomorphic
$G$-valued function on the cylinder. If we choose another solution
 of (4.3), say, $\psi_1$, it will have the form
$\psi_1(z)=\psi_0(z)\mu(z)$ where $\mu$ is holomorphic. Therefore, the
function $\nu_1(z)=\psi_1(z)^{-1}\psi_1(z+\tau)$ can be expressed as
follows:
$\nu_1(z)=\mu(z)^{-1}\nu_0(z)\mu_(z+\tau)$. This implies that the
conjugacy class of the element $(\tau, \nu_0)$
is independent on the choice of the solution $\psi_0$ of (4.3). Thus we
have canonically associated a conjugacy class of $\check G^{S^1}_h$ to any
equation of form (4.3). It is clear that every conjugacy class in
$C_{\tau}$ comes
from a certain equation. Thus we have established a one-to-one
correspondence between orbits of the action of $G^{\Sigma}$ in $\Cal
H^{\lambda}$ and conjugacy classes of $\check G^{S^1}_h$ in $C_{\tau}$.

Now the result that generic orbits of $G^{\Sigma}$ in $\Cal
H_{\lambda}$ are parametrized by points of the space $\Cal M$ can be deduced
from the classical theory of difference equations. This theory was
 developed in the beginnning of the twentieth century, by
G.D.Birkhoff, R.D.Carmichael, C.R.Adams, W.J.Trjitinsky and others \cite{3,17}.

 Consider the equation
$$
F(z+\tau)=A(z)F(z)\tag 4.4
$$
where $A(z)$ is an entire periodic $G$-valued
function with period 1.

\proclaim{Proposition 4.2}
For
almost every $A(z)$ (i.e. for $A(z)$ belonging to a Zariski open
subset in $G^{S^1}_h$) there exists a solution of (4.4) of the form
$$
F(z)=\Phi(z)\exp(Pz),\tag 4.5
$$
where $P\in\frak g$ and $\Phi$ is an entire $G$-valued function with
period 1.
\endproclaim

This proposition, for $G=SL_n$,
 can be deduced from the Fundamental Existence Theorem in the
theory of linear difference equations \cite{17}.

Clearly, Proposition 4.2 implies the classification result for generic
orbits.

Despite we did not use the notion of a vector bundle in this argument,
it was implicitly present in our considerations. Indeed,
E.Looijenga \cite{8} observed that the conjugacy classes of
$G^{S^1}_h$ which lie inside $C_{\tau}$
 are in one-to-one correspondence with
equivalence classes of holomorphic principal $G$-bundles over the
complex torus $\Sigma_{\tau}=\Bbb C/(\Bbb Z\oplus \tau\Bbb Z)$. This
correspondence is constructed as follows.

The complex torus $\Sigma_{\tau}$ can be obtained from
the annulus $\lbrace z\in\Bbb C/\Bbb Z|0\le \text{Im}z\le
\text{Im}\tau\rbrace$ by gluing together the boundary components according to
the rule $z\leftrightarrow z+\tau$. In order to define a holomorphic
bundle on the torus, it is enough to present a $G$-valued
holomorphic transition function $A(z)$ in a neighborhood of the
seam $\text{Im}z=0$. Therefore, we can naturally associate a holomorphic
$G$-bundle to every element $(\tau,g)\in C_w$ by setting $A(z)=g(z)$.
Now observe that equation (4.2) expresses exactly the fact that the
equivalence class of this bundle does not depend on
the choice of the element inside the conjugacy class, and that
different conjugacy classes give rise to non-equivalent bundles. It remains
to make sure that every holomorphic $G$-bundle over $\Sigma_{\tau}$ comes
from a certain conjugacy class in $C_{\tau}$. To see this, let us pick a
bundle $B$ over $\Sigma_{\tau}$, and pull it back to the cylinder $\Bbb
C/\Bbb Z$. Of course, the obtained bundle $\tilde B$ will be trivial.
Let us pick a global holomorphic section $\chi(z)$ of $\tilde B$. Then
$\chi(z+\tau)$ is another holomorphic section. Therefore,
$g(z)=\chi(z)^{-1}\chi(z+\tau)$ is a holomorphic function. Evidently,
the bundle $B$ is associated to the conjugacy class of $(\tau, g)$.

To conclude this section, let us discuss the coadjoint action of
$\text{Vir}(\Sigma)$ in the case when $\Sigma$ is a complex torus.
Obviously, this Lie algebra cannot be integrated to a Lie group --
this is impossible even in the one-dimensional case as long as we
work over $\Bbb C$. Therefore, we cannot define orbits of the
coadjoint action in the usual way. What we can do, though, is define
and fully describe holomorphic invariants of the coadjoint action.

Coadjoint orbits of the real
Virasoro algebra have been described in \cite {7,16}.
The smooth part of the
coadjoint representation can be interpreted as the space of Hill's
operators $\lambda\frac{d^2}{dx^2}+q(x)$ taking densities of weight
$-1/2$ to densities of weight $3/2$ on the circle. The coefficient
$\lambda$ is invariant under the coadjoint action, so we may
restrict this action to the hyperplane $\Cal H_{\lambda}$ of operators
with a fixed value of this coefficient. Orbits of the group
$\text{Diff}_+(S^1)$
lying in this hyperplane
for $\lambda\neq 0$ are labelled by conjugacy classes of the universal
covering of the group $SL_2(\Bbb R)$.

For the complex Virasoro algebra, orbits are not
defined since the algebra does not integrate to a Lie group. However, one can
study holomorphic invariants of the coadjoint action which carry the
same information as orbits. Repeating the above argument
with obvious modifications, one finds that besides $\lambda$
there is essentially a unique
holomorphic invariant of the coadjoint action  -- the monodromy of the
corresponding Hill's
operator. This invariant takes values in the space of conjugacy
classes of $SL_2(\Bbb C)$. Since almost every element in
$SL_2(\Bbb C)$ is diagonalizable, we may assume that the monodromy
invariant takes values in $\Bbb C^*/\sim $ where $\sim$ is the
equivalence relation that identifies $z$ with $z^{-1}$.
Any holomorphic invariant of
the coadjoint action will then be a function of the coefficient
$\lambda$ and
the monodromy invariant.

It is surprising that the same procedure works in two dimensions.
 Define a density of weight $\mu\in \Bbb C$ on the torus $\Sigma$ as a formal
expression $u(d\bar z)^{\mu}$ where $u$ is a smooth function on
$\Sigma$.
The action of (0,1)-vector fields on densities is given by
$$
v\frac{\partial}{\partial \bar z}\circ u(d\bar z)^{\mu}=
\left(v\frac{\partial u}{\partial \bar z}+\mu u
\frac{\partial v}{\partial \bar z}\right)(d\bar z)^{\mu}.\tag 4.6
$$
The smooth part of the coadjoint representation of $\text{Vir}(\Sigma)$ can
now be identified with the space of two-dimensional Hill's operators
$H=\lambda\frac{\partial^2}{\partial \bar z^2}+q(z,\bar z)$ taking
densities of weight $-1/2$ to densities of weight $3/2$  on the torus.

The monodromy of such an operator can be defined as follows. Assume
$\lambda\neq 0$. Consider
the differential equation
$$H u=0.\tag 4.7 $$
 This equation is equivalent to
the system
$$
\frac{\partial}{\partial \bar z}U+QU=0,\quad
Q=\left(\matrix 0 & 1\\ q & 0\endmatrix\right),\quad
U=\left(\matrix u\\ u^{\prime}\endmatrix\right),\quad u^{\prime}=
\frac{\partial u}{\partial \bar z}.\tag 4.8
$$
This system, as we know, defines a holomorphic $SL_2(\Bbb C)$-bundle
on the torus, which is almost always flat and unitary. Therefore, for
almost every $H$ defined is a point on $\Cal M$ corresponding
to this bundle. Let us call this point the monodromy of $H$.

\proclaim{Proposition 4.3}  The monodromy is a holomorphic invariant of the
coadjoint action of $\text{Vir}(\Sigma)$.
\endproclaim

\demo{Proof} The proof is by giving another definition of the monodromy.
 Let $\Cal S(H)$ be the set of all $s\in \Bbb C$ such that
there exist
 two linearly independent solutions $u_1$, $u_2$ of
(4.7) which are functions on the cylinder $\Bbb C/\Bbb Z$ and satisfy
the conditions
$$
u_1(z+\tau)=su_1(z),\quad u_2(z+\tau)=s^{-1}u_2(z).\tag 4.9
$$
$\Cal S(H)$ is non-empty whenever the bundle associated to
$H$ is flat and unitary, i.e. for almost every $H$.
Also, if $s\in \Cal S(H)$ then for any $n\in \Bbb Z$
$s^{(n)}=s\exp(2\pi in\tau)\in \Cal S(H)$. Indeed, if we
consider new solutions
$u_1^{(n)}=u_1\exp(2\pi inz)$, $u_2^{(n)}=u_2\exp(-2\pi inz)$ of (4.8),
they will satisfy (4.9) with $s$ replaced by $s^{(n)}$. Finally, if
$s\in \Cal S(H)$ then $s^{-1}\in \Cal S(H)$. Moreover, it is
easy to check that conversely, if $s_1,s_2\in \Cal S(H)$ then either
$s_2=s_1^{(n)}$ for some integer $n$ or $s_2^{-1}=s_1^{(n)}$.

Define $\hat s$ as the projection of $s$ into the space $\Bbb
C^*/\sim$ where $\sim$ is the equivalence relation that identifies
$s$ with $s^{(n)}$ for any $n$, and $s$ with $s^{-1}$. If $s\in\Cal
S(H)$ then $\hat s$ depends only on $H$. It is obvious from the
definition that $\hat s$ is a holomorphic invariant. It remains to
observe that $\Bbb C^*/\sim$ is isomorphic to $\Cal M$, and that $\hat s$ is
nothing else but the monodromy of $H$.
\enddemo

\proclaim{Remark 4.1}\rm
1. A little more extra work is needed to prove the following:

 Any holomorphic invariant of the coadjoint action of $\text{Vir}(\Sigma)$
is a function of $\lambda$ and the monodromy.

 This statement motivates a definition of coadjoint orbits for
$\text{Vir}(\Sigma)$ as level sets of the monodromy inside hyperplanes
$\lambda =\text{const}$ in the space of two-dimensional Hill's operators.
\endproclaim

\proclaim{Remark 4.2} \rm
Observe that the space of ``orbits''
for $\text{Vir}(\Sigma)$ is the same as
for for $SL_2(\Bbb C)^{\Sigma}$. A similar coincidence takes place also in
the one-dimensional theory, where it reflects the deep relationship
between the representation theories of Vir$_{\Bbb C}$ and $\widehat
{\frak sl_2}(\Bbb C)$.

\endproclaim

\heading
{\bf 5. Geometry of orbits.}
\endheading

Define an analogue of Kirillov-Kostant structure on orbits
of the coadjoint action of the group $G^{\Sigma}$. We will assume that
$\Sigma$ is a complex torus. The construction is the same as for
classical Lie groups.

Let $X_1,X_2$ be tangent vectors to an orbit $\Cal O$ in $E^*$ at a point
$f$. Then there exist elements $\tilde X_1,\tilde X_2\in\frak g^{\Sigma}$
such that
$$
\frac{d}{dt}(e^{t\tilde X_j}\circ f)=X_j.\tag 5.1
$$
Set
$$
K(X_1,X_2)=f([\tilde X_1,\tilde X_2])\tag 5.2
$$
Despite the liftings $\tilde X_1, \tilde X_2$ are not unique,
equation (5.2) presents a well defined antisymmetric bilinear
form on the tangent space to $\Cal O$ at $f$ for every $f\in \Cal O$.
Thus, $K$ is a holomorphic differential 2-form on $\Cal O$.

The following properties of $K$ are proved similarly to the classical
orbit method.

\proclaim{Proposition 5.1}

(i) $K$ is closed.

(ii) $K$ is nondegenerate.

(iii) $K$ is invariant under the action of $G^{\Sigma}$.
\endproclaim

\proclaim{Remark 5.1}\rm
 Unfortunately, we are unable to define a proper
counterpart of Kahler structure on the orbits. This appears to be
the main obstacle in attempts to construct projective
representations of $G^{\Sigma}$ via orbit method. Attempts to
construct representations algebraically encounter quite similar
difficulties arising from the inability to
generalize the notion of polarization for loop algebras
to $\frak g^{\Sigma}$.
\endproclaim

Let us describe the  topological structure of orbits. For this purpose
we need to describe the isotropy subgroup of a vector in the coadjoint
representation.

Recall the realization of $\Cal M$ as $(T\times T)/W$.
Let $m\in \Cal M$, and let $\tilde m\in T\times T$ be an inverse image of $m$.
Let $W_0\subset W$ be the stabilizer of $\tilde m$. We choose $\tilde
m$ so that $W_0$ be generated by a set of
Weyl reflections corresponding to the nodes of the Dynkin diagram of
$G$.
Also, pick an
inverse image $\hat m\in {\frak t}_{\Bbb C}$ of the point $\tilde m$ in the
complex Cartan subalgebra. Denote by $\text{Stab}(\hat m)$ the stabilizer of
$\hat m$ in the Weyl group. Obviously, we have $\text{Stab}(\hat m)\subset
W_0$. Also, it is easy to show that among the groups
$\text{Stab}(\hat m)$ for various inverse images $\hat m$ there is
a maximal one, i.e. one containing all the others. We denote this
group by $W^{\prime}$ and the corresponding inverse image by
$m^{\prime}$. The group
 $W_0^{\prime}$ coincides with the group of all $w\in W_0$ such that
$w\hat m-\hat
m=wq-q$, $q\in Q^{\vee}$, for an inverse image $\hat m$. It is also
generated by Weyl reflections.

 It follows from this definition that $W_0^{\prime}$ is a normal
subgroup in $W_0$. Denote the quotient $W_0/W_0^{\prime}$ by $F$.
Let $R$ be the centralizer of $m^{\prime}$ in $G$. It is
known that $R$ is a connected reductive subgroup of $G$ (Levi subgroup).
The Weyl group of $R$ is $W_0^{\prime}$. Obviously, $W_0$ normalizes
$R$, since it normalizes $T_{\Bbb C}$ and $W_0^{\prime}$. Therefore,
we can form a semidirect product $\tilde R=F\ltimes R$.

\proclaim{Proposition 5.2} The stabilizer of the element
$D=\lambda\frac{\partial}{\partial\bar z}+\lambda
m^{\prime}\in \Cal H_{\lambda}$ in
$G^{\Sigma}$ is isomorphic to $\tilde R$. Its intersection with the
subgroup of constant currents is equal to $R$.
\endproclaim

\proclaim{Remark 5.2}\rm
 A quite similar statement is true in the case of loop groups.
\endproclaim
\proclaim{Remark 5.3}\rm
 If the quotient $F$ is nontrivial, the stabilizer of $D$ is not
conjugate in $G^{\Sigma}$ to any subgroup of constant currents.
\endproclaim

\demo{Idea of proof} For the sake of brevity let us assume that
$G=SL_n$. Then the operator $D$ naturally acts in the space of $\Bbb
C^n$-valued functions on the torus.
If we let $\mu_k$, $1\le k\le n$ be the eigenvalues of $m^{\prime}$
as an operator in $\Bbb C^n$, and $v_k$, $1\le k\le n$
form the corresponding eigenbasis,
then the spectrum of $D$
is the union of $n$ lattices $\mu_k+L$,
and the corresponding eigenbasis of $D$ would be $v_k
\exp(2\pi i(px+qy))$ where $p,q\in \Bbb Z$ and $x,y$ are real coordinates
such that $z=x+\tau y$. In the generic case, when
$W_0$ is trivial and all the eigenvalues of $D$ are distinct, any element
$h$ from the stabilizer of $D$ in $G^{\Sigma}$ has to be diagonal in this
basis, which shows that the stabilizer of $D$ is
just $T_{\Bbb C}$. When $D$ has multiple eigenvalues, it is still true
that its eigenspaces are invariant under $h$. It allows one to
classify all possible $h$ and obtain the result of the theorem.
A similar argument works for an arbitrary $G$.
\enddemo

\proclaim{Corollary 5.3} The orbit $\Cal O_D$ of the vector $D$ is
isomorphic to $G^{\Sigma}/\tilde R$.
\endproclaim

\proclaim{Corollary 5.4} Let s=dim(R/[R,R]). Then
$H_2(\Cal O)=\Bbb Z^{s+2}\oplus \text{torsion}$.
\endproclaim

\demo{Proof} Pick a point $z\in \Sigma$ and decompose $G^{\Sigma}$ as
a smooth manifold as follows: $G^{\Sigma}=G_z^{\Sigma}\times G$,
$G_z^{\Sigma}$ being the group of all currents equal to the identity
at $z$. Since $G$ is 3-connected, we have
$H_2(G_z^{\Sigma})=H_2(G^{\Sigma})=\Bbb Z^2\oplus \text{torsion}$. Therefore,
$H_2(G^{\Sigma}/R)=H_2(G_z^{\Sigma}\times G/R)=H_2(G_z^{\Sigma})\oplus
H_2(G/R)$ (since $G/R$ is simply connected). The homology of the
flag space $G/R$ is known; in particular, $H_2(G/R)=\Bbb Z^s$.
Thus, $H_2(G^{\Sigma}/R)=\Bbb Z^{s+2}\oplus \text{torsion}$.
According to the above
proposition, the space $G^{\Sigma}/R$ is a finite covering of the
orbit $\Cal O_D$ with the fiber $F=\tilde R/R$. Since $F$
acts trivially on $R/[R,R]$, it also acts trivially on
$H_1(R)=H_2(G/R)$. Therefore, $H_2(\Cal O_D)=H_2(G^{\Sigma})$. $\blacksquare$
\enddemo

Let us construct a fundamental system of 2-cycles on $\Cal O_D$. First
of all, we have two independent cycles $C_x$ and $C_y$ that descend
from the group $G^{\Sigma}$. They are obtained by pushing forward the
generator of $H_2(G^{S^1})=\Bbb Z$ by the embeddings
$f_x,f_y:G^{S^1}\to G^{\Sigma}$ induced by the projections of
$\Sigma=S^1\times S^1$ onto the first and the second circle,
respectively.

Let us call a node $\nu$ of the Dynkin diagram of $G$ regular
if the corresponding Weyl reflection $w_{\nu}$ does not lie in
$W_0^{\prime}$. There are exactly $s$ regular
nodes: $\nu_1,...,\nu_s$.
To each regular node $\nu_j$ we can canonically
associate a 2-cycle $C_{\nu_j}$ in $G/R$, and transfer this cycle to
$\Cal O_D$. This transfer, however, is not canonical: it depends on
the choice of the element $m^{\prime}$. If we change this element to
another one, the cycle $C_{\nu_j}$ will be shifted by $p_jC_x+q_jC_y$
where $p_j,q_j$ are integers.

The cycles $C_x,C_y,C_{\nu_1},...,C_{\nu_s}$ constitute a basis in the
free part of $H_2(\Cal O_D)$.

At this point we need to assume a standard normalization of the invariant form
$<,>$. Namely, we pick the one in which the minimal length of a root
of $\frak g$ equals $\sqrt{2}$.

\proclaim{Proposition 5.5}
$$
\frac{1}{2\pi}\int_{C_x}K=\lambda,\quad
\frac{1}{2\pi}\int_{C_y}K=\lambda\tau,\quad
\frac{1}{2\pi}\int_{C_{\nu_j}}K=\lambda <h_{\nu_j},m^{\prime}>. \tag 5.3
$$
\endproclaim

\demo{Idea of proof} Since the cycles $C_x$ and $C_y$ lie inside
 the images of $G^{S^1}$, and the cycles $C_{\nu_j}$ lie inside the
subgroup
$G$
of constant currents, these identities follow from the corresponding
results for $G^{S^1}$.
\enddemo

Let us now define and classify integral orbits, by analogy with the
theory of loop groups \cite {13},\cite {5}. For loop groups, an integral orbit
is defined as an orbit on which the form $K/2\pi$ has integral
periods. A theorem in \cite {13, Section 4.5} states that an orbit
is integral if and only if there exists a circle bundle on it with
curvature $K$.

\proclaim{Definition 5.1} Let us say that an orbit $\Cal O_D\subset
\Cal H_{\lambda}$
is integral if for any $C\in H_2(\Cal O_D)$
$$
\frac{1}{2\pi}\int_CK\in L\tag 5.4
$$
\endproclaim

\proclaim{Proposition 5.6}

(i) An orbit $\Cal O_D$ is integral if and only if there
exists a holomorphic principal $\Sigma$-bundle $\Cal E$ over $\Cal O_D$ with
a connection $\theta$ whose curvature is $K$.

(ii) If $\Cal O_D$ is an integral orbit and $\Cal E$ is the corresponding
principal bundle then the action of $G^{\Sigma}$ on $\Cal O_D$
can be uniquely lifted to an action of $\hat G^{\Sigma}$ on $\Cal E$
preserving the connection $\theta$.
\endproclaim

The proof is similar to that for loop groups.

Identities (5.3) allow us to classify integral orbits.

Denote by $\Cal O(\lambda,m)$ the orbit in $\Cal H_{\lambda}$
corresponding to the point $m\in \Cal M$. We assume that $\lambda\ne
0$. Let us call $\lambda$ the
level and $m$ the weight of the orbit, by analogy with loop
groups.

We also assume that $\tau$ is generic, i.e. that $\Sigma$ has exactly
two holomorphic automorphisms modulo translations.

Denote by $Z$ the center of $G$. Let $\lambda\in \Bbb C^*$, $m\in \Cal
M$, and let $\tilde m\in T\times T$ be a lifting of $m$.

\proclaim{Proposition 5.7} An orbit $\Cal O(\lambda,m)$ is integral if and
only if $\lambda$ is an integer, and $\lambda\tilde m\in Z\times
Z\subset T\times T$.
Thus, the number of integral orbits at each level is finite.
\endproclaim

This statement follows immediately from the two previous propositions.

The number of integral orbits at each level can be easily calculated
for any particular group. For instance, if $G=SL_2$ then it is equal
to $2\lambda^2+2$.

\centerline{\bf Acknowledgements.}
\vskip .1in

We would like to thank R.Beals and H.Garland for interesting
discussions. We are grateful to R.Beals for showing us an elementary
analytic proof of Theorem 3.3 for the case of a surface of genus one
and $G=SL_n$. The work of I.F. was supported by NSF grant DMS-8906772.

\end